\documentclass[12pt,a4paper]{article}

\usepackage{amsmath,amssymb}
\usepackage{graphicx} 
\PassOptionsToPackage{hyphens}{url}\usepackage[pdfusetitle,pdfauthor={RC,BSM,KD,RRS,RL},colorlinks=true,linkcolor=teal,citecolor=magenta]{hyperref}
\usepackage[table,dvipsnames]{xcolor}
\usepackage{array}
\usepackage{microtype}
\usepackage[margin={2.54cm, 2.54cm}]{geometry}
\usepackage[T1]{fontenc}
\usepackage[sort&compress]{natbib}
\setcitestyle{square,numbers,comma}

\usepackage{colortbl}

\definecolor{mygreen}{RGB}{104,170,131}

\definecolor{green0}{RGB}{255,255,255} 
\colorlet{green1}{mygreen!15!white}
\colorlet{green2}{mygreen!30!white}
\colorlet{green3}{mygreen!45!white}
\colorlet{green4}{mygreen!60!white}
\colorlet{green5}{mygreen!75!white}
\colorlet{green6}{mygreen}

\newcommand{\ccv}[1]{%
  \begingroup
  \def\val{#1}%
  \ifnum\val<5 \cellcolor{green0}%
  \else\ifnum\val<10 \cellcolor{green1}%
  \else\ifnum\val<15 \cellcolor{green2}%
  \else\ifnum\val<20 \cellcolor{green3}%
  \else\ifnum\val<25 \cellcolor{green4}%
  \else\ifnum\val<30 \cellcolor{green5}%
  \else \cellcolor{green6}%
  \fi\fi\fi\fi\fi\fi
  \val\%%
  \endgroup
}

\newcounter{principle}
\renewcommand{\theprinciple}{\ifnum\value{principle}<10 0\fi\arabic{principle}}
\newcommand{\principle}{\refstepcounter{principle}\theprinciple}

\newcommand{\email}[1]{\thanks{\href{mailto:#1}{#1}}}

\title{Exploring Quantum Responsible Innovation efforts in Canada and the world}

\author{
    Ria Chakraborty,%
    $^{1}$%
    \and
    Bruna S. de Mendonça,%
    $^{2}$%
    \email{bruna.mendonca@cmc.ca}
    \and
    Katya Driscoll,%
    $^{1}$%
    \and
    Rodolfo R. Soldati,%
    $^{1,3,4}$%
    \email{rreissoldati@uwaterloo.ca}
    \and
    Ray Laflamme%
    $^{1,3,4}$%
}

\date{\normalsize
    $^{1}$Institute for Quantum Computing, University of Waterloo, Waterloo, ON, N2L 3G1, Canada.
    \\
    $^{2}$CMC Microsystems, Sherbrooke, QC J1K1B8, Canada.
    \\
    $^{3}$Perimeter Institute for Theoretical Physics, Waterloo, ON, N2L 2Y5, Canada.
    \\
    $^{4}$Department of Physics and Astronomy, University of Waterloo, Waterloo, ON, N2L 3G1, Canada.
}

\begin{document}

\maketitle

\begin{center}\small
    \textbf{Short summary:} We assess global quantum technologies efforts in the
    Responsible Innovation framework and how Canada leads or lags in development
    and regulations.
\end{center}

\begin{abstract}
    The global landscape for quantum technologies (QTs) is rapidly changing, and
    proper understanding of their impact and subsequent regulations need to
    match this pace. A Responsible Innovation (RI) approach and guiding
    principles have been proposed to accompany this development. We examine
    practical efforts globally and in Canada, from industry to research to
    governments, and analyze the current status of quantum technological
    advances under the RI framework. We analyze and compare what is being done
    internationally, identify gaps in the Canadian strategy, propose initiatives
    to fill those gaps, and highlight areas where Canada is leading or where
    more work is needed.
\end{abstract}

The widespread real-world impact that quantum devices will have in the near
future demands attention from various sectors of society, not only those
interested in research and commercialization. In response to this, international
efforts have started to address the needs and consequences of the use of quantum
technologies \cite{kung2021}.

An integral part of the allocation of quantum technologies in society is
consideration of the ethical implications of their use, operation, and
manufacture. Do countries responsible for the research and implementation of
such devices take these implications into account  as they lay out their
roadmaps for development? How does Canada fit in this larger picture?

We review efforts from governments, private initiatives, research groups, and
grants to answer these questions in the frame of responsible
innovation~\cite{kop2024,burch2024}. Whenever possible, we highlight verifiable
output displaying the efforts: events, technical papers, policy documents, etc.
We use \citet{kop2024} to determine the grounds for comparison, and survey
the relative frequency with which each of the principles of responsibility are
addressed within a National Quantum Strategy (NQS).

We search and collect in NQS documents the keywords defined in
Table~\ref{tab-keywords}. This is followed by verifying, within context, whether
the sentence containing the keyword addresses the corresponding Principle.

The accounted quotes may include similar words, e.g.~``ethics'' and ``ethical.''
Finally, the mentions to each Principle are counted and compiled in
Table~\ref{tab-quotes}.\footnote{We follow the methods used to compile Figure 2
in Lesher et al.~\cite{davidgierten2022}.} We use the CIFAR
report~\cite{kung2021} to decide what document is an NQS, and we include updated
references by Qureca~\cite{qureca2024} and based on other web searches. We do
not attempt to define new criteria.

We refer to each of the ten Principles in an abbreviated manner: P\ref{01}, for
example, refers to Principle 1 (Information security). Additional remarks on the
choice of NQS documents are discussed in the next section.

\begin{table}
    \resizebox{\columnwidth}{!}{\footnotesize
    \begin{tabular}{r >{\raggedright}p{2.4cm}|>{\raggedright}p{5cm}|>{\raggedright\arraybackslash}p{5cm}}
    \hline
    \rowcolor{gray!10}
    {} & Principle & Description \cite{kop2024} & Keywords
    \\
    \hline
    \principle\label{01} &
    Information Security &
    Make information security an integral part of QT. &
    Security, cybersecurity, threat, attack 
    \\
    \rowcolor{gray!10}
    \principle\label{02} &
    Dual Use &
    Proactively anticipate the malicious use of quantum applications. &
    Dual use, usage, malicious, hazard, warfare, unintended 
    \\
    \principle\label{03} &
    Quantum Race &
    Seek international collaboration based on shared values. &
    Race, international, collaboration 
    \\
    \rowcolor{gray!10}
    \principle\label{04} &
    Quantum Gap &
    Consider our planet as the sociotechnical environment in which QT should function. &
    Gap, quantum divide, global, world, unequal, disproportionate, fair 
    \\
    \principle\label{05} &
    Intellectual Property &
    Incentivise innovation while being as open as possible and as closed as necessary. &
    IP, Intellectual Property, open source 
    \\
    \rowcolor{gray!10}
    \principle\label{06} &
    Inclusion &
    Pursue diverse R\&D communities in terms of disciplines and people. &
    Inclusion, DEI or EDI, equity, diversity 
    \\
    \principle\label{07} &
    Societal Relevance &
    Link quantum R\&D explicitly to desirable societal goals. &
    Social, benefit
    \\
    \rowcolor{gray!10}
    \principle\label{08} &
    Complementary Innovation &
    Actively stimulate sustainable, cross-disciplinary innovation. &
    Innovation, adaptation, adoption, progress 
    \\
    \principle\label{09} &
    Responsibility &
    Create an ecosystem to learn about the possible uses and consequences of QT applications. &
    Responsible, ethics, consequences, risk 
    \\
    \rowcolor{gray!10}
    \principle\label{10} &
    Education and Dialogue &
    Facilitate dialogues with stakeholders to better envision the future of QT. &
    Education, dialogue, outreach, science communication, general public, engage \\
    \hline
    \end{tabular}
}
    \caption{List of principles, a short description taken directly from \cite{kop2024}, and our choices of keywords for the keyword search.}
    \label{tab-keywords}
\end{table}

\section{International}

We analyze the NQS documents listed in Table~\ref{tab-quotes} and display the
keyword search results there.

We see that all NQS documents roughly follow the same pattern, which reflects
the context and time when these documents were crafted, with the most pressing
issues of information security (P\ref{01}) and race to technological edge
(P\ref{03}) heading the counts, and with issues of innovation (P\ref{08})
following. Occupying a negligible portion of each NQS, we have Dual Use
(P\ref{02}), Quantum Gap (P\ref{04}), and IP (P\ref{05}). A possible explanation
is that both Dual Use and IP regulations may be covered in more specialized
documents, when contrasted with Information Security due to its national
security implications. Furthermore, while the absence of mentions of the Quantum
Gap may be explained by the country-specific scope of these documents, it is an
example of terminology that has only been introduced recently.

We expect that growing attention to societal problems will be given when QT
reaches commercial application with practical use cases. These results suggest
that the Principles of Quantum Gap, Inclusion, and Education (P\ref{04},
P\ref{06}, P\ref{10}), which can inform what problems are of Societal Relevance
(P\ref{07}), are lagging behind.

The development of QT happens in an effervescent ecosystem, as demonstrated by
the website for the International Year of Quantum (IYQ) \cite{unitednations},
but a lot still needs to be addressed. The Quantum Economy Blueprint
\cite{quantumeconomynetwork2024}, a report guiding responsible development,
raises the problem of the Quantum Divide (P\ref{04}) between developed and
Majority World countries as one that needs addressing. The IYQ map gives a
glimpse of this: the number of activities in North America and Europe largely
outnumber the ones in Latin America, Africa, and Asia. As we will argue in the
following sections, research and an active stance must be taken on this, and
with regards to other Principles that received less attention in the NQS
documents, as we move forward. Furthermore, in an agenda to be fulfilled by
2030, the United Nations define Sustainable Development Goals (SDGs), and it has
been argued that QTs have the potential to be used to address them (P\ref{07})
\cite{dimeglio2022,openquantuminstitute2024,openquantuminstitute2024a}, and
policies should reflect that \cite{oecd2025}.

Before focusing on Canada, we point out that an integral part of the
international efforts is led by specialists and address a large set of problems
that can be classified under the RI Principles. Some examples are QT activities
in the G77 plus China (P\ref{04}), a study commissioned by the Quantum Delta
program \cite{awoagency2024}. Research groups are starting to tackle the field
of ``quantum humanities'' (P\ref{07}--\ref{09}), with examples in Europe, Asia,
and North America
\cite{loos2024,dongen,malaysiaquantuminformationinitiative,quantumethicsproject,arrow2025},
and to aid in this, quantum experts must have a strong grasp of QT (P\ref{10})
\cite{ruane2025}. Research on resource costs for QT, and how QT can address
climate change are also integrated under socially relevant problems
(P\ref{07})and are being investigated \cite{berger2021,auffeves2022}.
Ultimately, QT must not be considered in isolation, and its interaction with
other important emerging technologies is paramount, in particular artificial
intelligence (P\ref{08}) \cite{acampora2025,giovanniacampora2025}.

\begin{table}
\resizebox{\columnwidth}{!}{
    \begin{tabular}{p{0.24\linewidth}|*{10}{r}}
    \hline
    \rowcolor{gray!10}\raggedright
    National Quantum Strategies \cite{kung2021,qureca2024} & \multicolumn{10}{c}{Principles \cite{kop2024}} \\\rowcolor{gray!10}
    {} & \ref{01} & \ref{02} & \ref{03} & \ref{04} & \ref{05} & \ref{06} & \ref{07} & \ref{08} & \ref{09} & \ref{10} \\
    \hline
    Australia \cite{ausNQS}         & \ccv{10} & \ccv{0} & \ccv{27}
                                    & \ccv{0} & \ccv{4} & \ccv{12}
                                    & \ccv{10} & \ccv{6} & \ccv{16}
                                    & \ccv{14} \\\rowcolor{gray!10}
    Canada \cite{canNQS}            & \ccv{11} & \ccv{1} & \ccv{40}
                                    & \ccv{1} & \ccv{2} & \ccv{5}
                                    & \ccv{14} & \ccv{9} & \ccv{9}
                                    & \ccv{9} \\
    Denmark \cite{danNQS,danNQSa}   & \ccv{22} & \ccv{2} & \ccv{26}
                                    & \ccv{1} & \ccv{3} & \ccv{2}
                                    & \ccv{12} & \ccv{18} & \ccv{9}
                                    & \ccv{4} \\\rowcolor{gray!10}
    Ireland \cite{ireNQS}           & \ccv{23} & \ccv{1} & \ccv{17}
                                    & \ccv{1} & \ccv{2} & \ccv{6}
                                    & \ccv{13} & \ccv{24} & \ccv{7}
                                    & \ccv{5} \\
    Netherlands \cite{nlNQS}        & \ccv{19} & \ccv{0} & \ccv{15}
                                    & \ccv{1} & \ccv{0} & \ccv{0}
                                    & \ccv{29} & \ccv{16} & \ccv{7}
                                    & \ccv{12} \\\rowcolor{gray!10}
    South Africa \cite{saNQS}       & \ccv{30} & \ccv{0} & \ccv{22}
                                    & \ccv{0} & \ccv{8} & \ccv{3}
                                    & \ccv{5} & \ccv{15} & \ccv{0}
                                    & \ccv{17} \\
    South Korea \cite{korNQS}       & \ccv{24} & \ccv{1} & \ccv{19}
                                    & \ccv{2} & \ccv{1} & \ccv{6}
                                    & \ccv{13} & \ccv{25} & \ccv{3}
                                    & \ccv{6} \\\rowcolor{gray!10}
    United Kingdom \cite{ukNQS}     & \ccv{16} & \ccv{0} & \ccv{20}
                                    & \ccv{0} & \ccv{4} & \ccv{4}
                                    & \ccv{15} & \ccv{22} & \ccv{14}
                                    & \ccv{4} \\
    United States \cite{usNQS}      & \ccv{32} & \ccv{2} & \ccv{19}
                                    & \ccv{0} & \ccv{6} & \ccv{0}
                                    & \ccv{6} & \ccv{14} & \ccv{5}
                                    & \ccv{16} \\
    \hline
    \end{tabular}
}
    \caption{Assessment of the presence of Principles in National Quantum
    Strategies. Numbers indicate how many quotes address each Principle, shown
    as a percentage of the total number of quotes in that document. These
    documents were chosen by referring to the countries with NQS ready or in
    development. We disregard NQS documents that are not in English or not
    publicly available.}
    \label{tab-quotes}
\end{table}

\section{Canada}

The Canadian NQS aims to establish Canada as a world leader on the quantum
development stage, to boost national cybersecurity with post-quantum
cryptography incentives, and to support the adoption of quantum sensing
technologies \cite{canNQS}.

Although all the principles outlined in the RI framework \ref{tab-keywords} are
discussed in the NQS, our analysis (see Table~\ref{tab-quotes}) revealed that
Canada's initiatives align primarily with Quantum Race (P\ref{03}) and Societal
Relevance (P\ref{07}). In particular, Canada placed a stronger emphasis on
P\ref{03} than any other country surveyed, demonstrating a particular focus on
international collaboration for successful quantum development. In contrast,
less attention was given to Inclusion (P\ref{06}), Complementary Innovation
(P\ref{08}) and Education and Dialogue (P\ref{10}) than in peer countries. To
align more closely with an RI framework, greater attention must be paid to these
principles. 

Bolstering one Principle often reinforces others. For example, increased
commitment to Inclusion (P\ref{06}) and Education (P\ref{10}) is needed to
properly bring Societal Relevance to light (P\ref{07}). After all, it is
difficult to identify the most relevant societal problems for quantum technology
without first strengthening diversity and engagement in the workforce and the
general public space. At the same time, it is important to consider the
potential trade-offs from Principle interactions. For one, increasing open
access (P\ref{05}) while maintaining a competitive edge in the quantum race
(P\ref{03}) and preserving national cybersecurity (P\ref{01}) poses a real
strategic challenge. It is important to keep this balancing act in mind when
making these changes.

Although still in its infancy, the Canadian responsible quantum innovation
landscape has already begun to develop, with some initiatives making use of
government funding. One such organization is the Centre for Responsible Quantum
Innovation and Technology (CRQIT) in British Columbia. With the principal goal
of nurturing responsible quantum innovation on a national and international
level (P\ref{02}, P\ref{03}, P\ref{09}), the CRQIT's efforts have a broad span.
This includes advisory work on EDI policies in Canada's quantum hubs and
engagement with UNESCO \cite{quantumalgorithmsinstitute}.

A project sharing similar goals is underway at the Institut Quantique. This
work, created with the realities of research and entrepreneurial development in
mind, aims to create tools to encourage those in the quantum ecosystem to
analyze the potential societal impacts of their work throughout its process,
rather than after the fact \cite{universityofsherbrooke2024} (P\ref{02},
P\ref{07}, P\ref{09}). Another important facet of the Canadian RI landscape has
been enabling broader participation in technological development. For instance,
Open Quantum Design is a non-profit organization in Waterloo creating an
open-source quantum computer using trapped ions \cite{openquantumdesigna}
(P\ref{09}, P\ref{10}). 

Finally, Canadians are also contributing to the growing international
conversation on responsible quantum innovation. The Centre for International
Governance Innovation (CIGI), based in Waterloo, has produced a number of
publications exploring Canada's quantum ecosystem through the lens of policy
\cite{michaelpamurphy2024,michaelpamurphy2024a,michaelpamurphy2025,kristencsenkey2025}.
In 2025, CIGI organized the Think7 (T7) process as part of Canada's G7
presidency, bringing together experts from around the world to develop policy
briefs that will inform G7 decision-making (P\ref{03}) \cite{think7,think7a}.

\section{Future prospects}

Canada's evolving quantum ecosystem offers a timely opportunity to align its
national initiatives with RI Principles. This section outlines recommendations
for Canada to strengthen its strategic position while ensuring ethical, secure,
and sustainable development of quantum technologies.

To begin, to guarantee a science-based adoption system, the government should
increase efforts to educate policymakers and companies about quantum
technologies (P\ref{10}). Well-defined standards assessing the development,
security, and effectiveness of quantum systems can only be properly defined if
those creating them have sufficient technical understanding. For instance, the
government could hire physicists to deliver workshops to help these
professionals build their quantum knowledge \cite{seskir2023}. Conversely, it is
also crucial for researchers to consider the social consequences of their work,
and one way to do this is by learning about policy. This will increase
engagement between the public, private, and academic sectors, clarify security
requirements, and increase trust in government regulation (P\ref{08},
P\ref{09}).

Partnerships with allies should continue to be nurtured to share the
responsibilities of advancing quantum technology (P\ref{03}). Moreover, Canada
has the opportunity to improve its position as a leader in quantum development
by actively pursuing inclusive relations with developing nations and pushing for
a shared-values system. Building such associations gives Canada leverage in
negotiations, provides a competitive edge in the Quantum Race (P\ref{03}), and
directly lessens the Quantum Divide (P\ref{04}). Throughout this work, it is
also important to remain aware of research security risks (P\ref{01}).

In our analysis, we found that most NQSs placed less emphasis on social
Principles like P\ref{06}. In Canada's case, although the importance of social
science research for quantum was discussed, no funding was assigned to it. We
suggest that a portion of NQS funding should be allocated to social science
research to connect the technical advancement of the quantum ecosystem to
ethical and regulatory considerations \cite{chakraborty2025}.

It is also important to encourage private companies to share insights from
real-world testing, thus fostering a more collaborative, open, and responsive
quantum ecosystem (P\ref{05}). Moreover, both the public and private sectors
should map out supply chains for future quantum technologies and take steps to
protect critical hardware from potential threats. This work protects quantum
infrastructure (P\ref{01}) and pushes nations to navigate dependencies and
maintain competitive resilience in the global quantum landscape (P\ref{03}).

Although this article focuses on Canada, an interesting generalization would be
to analyze the differences in focus between the Global North and South. We also
hope that the discussions in this paper will motivate physicists to become
involved in ethics and policy, thus ensuring the development of directives that
are both compliant with RI and scientifically accurate.

\section*{Acknowledgements}

The authors acknowledge comments from Zeki Seskir. R. L., R. C., K. D., and R. R. S. thank Mike and Ophelia Lazaridis for funding. R. C., B. S. M., K. D., and R. R. S. gratefully acknowledge the essential contributions of Ray Laflamme, who sadly passed away before the publication of this work. Research at Perimeter Institute is supported in part by the Government of Canada through the Department of Innovation, Science and Industry Canada and by the Province of Ontario through the Ministry of Colleges and Universities.


\begin{thebibliography}{43}
\providecommand{\natexlab}[1]{#1}
\providecommand{\url}[1]{\texttt{#1}}
\expandafter\ifx\csname urlstyle\endcsname\relax
  \providecommand{\doi}[1]{doi: #1}\else
  \providecommand{\doi}{doi: \begingroup \urlstyle{rm}\Url}\fi

\bibitem[Kung and Fancy(2021)]{kung2021}
J. Kung and M. Fancy, A Quantum Revolution: Report on Global Policies for Quantum Technology, CIFAR, Apr. 2021. \url{https://cifar.ca/wp-content/uploads/2021/05/QuantumReport-EN-May2021.pdf} (accessed 2025-05-13).

\bibitem[Kop et~al.(2024)Kop, Aboy, Jong, Gasser, Minssen, Cohen, Brongersma, Quintel, Floridi, and Laflamme]{kop2024}
M. Kop et al., Ten Principles for Responsible Quantum Innovation, Quantum Sci. Technol., 9, 035013 (Apr. 2024).

\bibitem[Burch et~al.(2024)Burch, {Finlay-Smits Susanna}, and Roberson]{burch2024}
K. A. Burch, S. Finlay-Smits, and T. Roberson, Responsible Innovation Is Not Comfortable: A Call for Grounded, Embodied Reflexivity When Doing RI, J. Responsible Innov., 11, 2427429 (Dec. 2024).

\bibitem[{David Gierten} and {Molly Lesher}(2022)]{davidgierten2022}
D. Gierten and M. Lesher, Assessing National Digital Strategies and Their Governance, OECD, May 2022. \url{https://www.oecd.org/en/publications/assessing-national-digital-strategies-and-their-governance_baffceca-en.html} (accessed 2025-05-06).

\bibitem[{Qureca}(2024)]{qureca2024}
Qureca, Quantum Initiatives Worldwide 2025, Apr. 2024. \url{https://www.qureca.com/quantum-initiatives-worldwide/} (accessed 2025-05-08).

\bibitem[{United Nations}()]{unitednations}
United Nations, IYQ 2025, \url{https://quantum2025.org/} (accessed 2025-04-22).

\bibitem[{Quantum Economy Network}(2024)]{quantumeconomynetwork2024}
Quantum Economy Network, Quantum Economy Blueprint, Jan. 2024. \url{https://www.weforum.org/publications/quantum-economy-blueprint/} (accessed 2025-04-22).

\bibitem[Di~Meglio et~al.(2022)Di~Meglio, Doser, Frisch, Grabowska, Pierini, and Vallecorsa]{dimeglio2022}
A. Di Meglio et al., CERN Quantum Technology Initiative Strategy and Roadmap, Zenodo, Jan. 2022. \url{https://zenodo.org/record/5553774} (accessed 2025-06-12).

\bibitem[{Open Quantum Institute}(2024{\natexlab{a}})]{openquantuminstitute2024}
Open Quantum Institute, Intelligence Report on Quantum Diplomacy for the Sustainable Development Goals (SDGs), CERN, Oct. 2024. \url{https://open-quantum-institute.cern/wp-content/uploads/2025/03/GESDA_OQI_Intelligence-Report-2024_Final.pdf}.

\bibitem[{Open Quantum Institute}(2024{\natexlab{b}})]{openquantuminstitute2024a}
Open Quantum Institute, SDG Use Cases. White Paper, CERN, Oct. 2024. \url{https://open-quantum-institute.cern/wp-content/uploads/2024/12/OQI_WhitePaper2024.pdf}.

\bibitem[{OECD}(2025)]{oecd2025}
OECD, A Quantum Technologies Policy Primer. Working Paper, Jan. 2025. \url{https://www.oecd.org/content/dam/oecd/en/publications/reports/2025/01/a-quantum-technologies-policy-primer_bdac5544/fd1153c3-en.pdf}.

\bibitem[{AWO Agency}(2024)]{awoagency2024}
AWO Agency, Quantum Computing and the G-77, Apr. 2024. \url{https://www.ivir.nl/publications/quantum-computing-and-the-g-77/quantum-computing-and-the-g-77/} (accessed 2025-04-22).

\bibitem[Loos et~al.(2024)Loos, Bickert, Dotzel, Tutschku, and Kaiser]{loos2024}
S. Loos et al., Potentials and Needs of the Quantum Computing-Ecosystem, Nov. 2024. \url{https://publica.fraunhofer.de/entities/publication/6f66f4ec-725f-46e5-bd4f-d1e69af0b375} (accessed 2025-06-17).

\bibitem[van Dongen and Weiglhofer()]{dongen}
E. van Dongen and H. Weiglhofer, Quantum Humanities Network EN, \url{https://www.uibk.ac.at/projects/iqel/team/quantum-humanities-network/index.html.en} (accessed 2025-04-22).

\bibitem[{Malaysia Quantum Information Initiative}()]{malaysiaquantuminformationinitiative}
Malaysia Quantum Information Initiative, MyQI - About, \url{https://www.myqi.my/about} (accessed 2025-06-13).

\bibitem[{Quantum Ethics Project}()]{quantumethicsproject}
Quantum Ethics Project, \url{https://www.quantumethicsproject.org} (accessed 2025-04-21).

\bibitem[Arrow(2025)]{arrow2025}
J. Arrow, Don’t Believe the Hype — Quantum Tech Can’t yet Solve Real-World Problems, Nature, 640, 572 (Apr. 2025).

\bibitem[Ruane et~al.(2025)Ruane, Kiesow, Galatsanos, Dukatz, Blomquist, and Shukla]{ruane2025}
J. Ruane et al. Quantum Index Report 2025. June 2025. \url{https://arxiv.org/abs/2506.04259} (accessed 2025-06-07).

\bibitem[Berger et~al.(2021)Berger, Paolo, Forrest, Hadfield, Sawaya, St{\k e}ch{\l}y, and Thibault]{berger2021}
C. Berger et al., Quantum Technologies for Climate Change: Preliminary Assessment, June 2021. \url{https://arxiv.org/abs/2107.05362} (accessed 2025-04-22).

\bibitem[Auff{\`e}ves(2022)]{auffeves2022}
A.~Auff{\`e}ves.
A. Auffèves, Quantum Technologies Need a Quantum Energy Initiative, PRX Quantum, 3, 020101 (June 2022).

\bibitem[Acampora et~al.(2025)Acampora, Ambainis, Ares, Banchi, Bhardwaj, Binosi, Briggs, Calarco, Dunjko, Eisert, Ezratty, Erker, Fedele, {Gil-Fuster}, G{\"a}rttner, Granath, Heyl, Kerenidis, Klusch, Kockum, Kueng, Krenn, L{\"a}ssig, Macaluso, Maniscalco, Marquardt, Michielsen, {Mu{\~n}oz-Gil}, M{\"u}ssig, Nautrup, van Nieuwenburg, Orus, Schmiedmayer, Schmitt, Slusallek, Vicentini, Weitenberg, and Wilhelm]{acampora2025}
G. Acampora et al. Quantum Computing and Artificial Intelligence: Status and Perspectives. May 2025. \url{https://arxiv.org/abs/2505.23860} (accessed 2025-06-02).

\bibitem[{Giovanni Acampora} et~al.(2025){Giovanni Acampora}, {Andris Ambainis}, {Natalia Ares}, {Leonardo Banchi}, {Pallavi Bhardwaj}, {Daniele Binosi}, {G. Andrew D. Briggs}, {Vedran Dunjko}, {Jens Eisert}, {Paul Erker}, {Olivier Ezratty}, {Federico Fedele}, {Elies Gil-Fuster}, {Mats Granath}, {Martin G{\"a}rttner}, {Markus Heyl}, {Iordanis Kerenidis}, {Matthias Klusch}, {Anton Frisk Kockum}, {Mario Krenn}, {Richard Kueng}, {J{\"o}rg L{\"a}ssig}, {Antonio Macaluso}, {Sabrina Maniscalco}, {Florian Marquardt}, {Kristel Michielsen}, {Daniel M{\"u}ssig}, {Gorka Mu{\~n}oz-Gil}, {Hendrik Poulsen Nautrup}, {Evert van Nieuwenburg}, {Roman Orus}, {J{\"o}rg Schmiedmayer}, {Markus Schmitt}, {Philipp Slusallek}, {Filippo Vicentini}, {Christof Weitenberg}, and {Frank K. Wilhelm}]{giovanniacampora2025}
G. Acampora et al. Feedback to Quantum Strategy of the EU. June 2025. \url{https://ec.europa.eu/info/law/better-regulation/have-your-say/initiatives/14675-Quantum-Strategy-of-the-EU/F3563275_en} (accessed 2025-06-13).

\bibitem[{Australian Government Department of Industry, Science and Resources}(2023)]{ausNQS}
Australian Government Department of Industry, Science and Resources, National Quantum Strategy, 2023. \url{https://www.industry.gov.au/sites/default/files/2023-05/national-quantum-strategy.pdf} (accessed 2025-05-06).

\bibitem[{Innovation, Science and Economic Development Canada}(2022)]{canNQS}
Innovation, Science, Economic Development Canada, Canada’s National Quantum Strategy, Government of Canada, 2022. \url{https://ised-isde.canada.ca/site/national-quantum-strategy/sites/default/files/attachments/2022/NQS-SQN-eng.pdf} (accessed 2025-05-05).

\bibitem[{Danish Government}(2023{\natexlab{a}})]{danNQS}
Danish Government, Denmark’s Strategy for Quantum Technology Part 1, Ministry of Higher Education and Science, June 2023. ISBN: 978-87-93807-66-2. https://ufm.dk/en/publications/2023/strategy-for-quantum-technology-part-1-2013-world-class-research-and-innovation.

\bibitem[{Danish Government}(2023{\natexlab{b}})]{danNQSa}
{Danish Government}.
Danish Government, Denmark’s National Strategy for Quantum Technology Part 2, Ministry of Industry, Business and Financial Affairs, Sept. 2023. ISBN: 978-87-94224-57-4. https://www.eng.em.dk/publications/2023/national-strategy-for-quantum-technology.

\bibitem[{Department of Further and Higher Education, Research, Innovation and Science}(2023)]{ireNQS}
Department of Further and Higher Education, Research, Innovation and Science, Quantum 2030 - A National Quantum Technologies Strategy for Ireland, Nov. 2023. \url{https://www.gov.ie/en/department-of-further-and-higher-education-research-innovation-and-science/publications/quantum-2030-a-national-quantum-technologies-strategy-for-ireland/} (accessed 2025-05-14).

\bibitem[{Quantum Delta}(2019)]{nlNQS}
Quantum Delta, National Agenda on Quantum Technology: The Netherlands as an International Centre for Quantum Technology, Sept. 2019. \url{https://qutech.nl/2019/09/16/national-agenda-on-quantum-technology-the-netherlands-as-an-international-centre-for-quantum-technology/} (accessed 2025-05-14).

\bibitem[{South African Quantum Technology Initiative}(2021)]{saNQS}
South African Quantum Technology Initiative, SAQUTI Framework Resources, Jan. 2021. \url{https://saquti.org/resources/} (accessed 2025-05-14).

\bibitem[{Ministry of Science and ICT}()]{korNQS}
Ministry of Science and ICT, Korea’s National Quantum Strategy, \url{https://www.msit.go.kr/eng/bbs/view.do?sCode=eng&mId=10&mPid=9&pageIndex=&bbsSeqNo=46&nttSeqNo=18&searchOpt=ALL&searchTxt=} (accessed 2025-05-08).

\bibitem[{Department for Science, Innovation and Technology}(2023)]{ukNQS}
Department for Science, Innovation and Technology, UK National Quantum Strategy, Mar. 2023. \url{https://www.gov.uk/government/publications/national-quantum-strategy} (accessed 2025-05-13).

\bibitem[{US National Quantum Initiative}()]{usNQS}
US National Quantum Initiative, US’s National Quantum Strategy, \url{https://www.quantum.gov/strategy/} (accessed 2025-05-15).

\bibitem[{Quantum Algorithms Institute}()]{quantumalgorithmsinstitute}
Quantum Algorithms Institute. Centre for Responsible Quantum Innovation and Technology. \url{https://www.qai.ca/centre-for-responsible-quantum-innovation-and-technology} (accessed 2025-05-06).

\bibitem[{University of Sherbrooke}(2024)]{universityofsherbrooke2024}
University of Sherbrooke. Rapport Annuel 2023–2024. 2024. \url{https://www.usherbrooke.ca/iq/fileadmin/sites/iq/uploads/iQ_RA_2023-2024_vf.pdf.} (accessed 2025-05-06).

\bibitem[{Open Quantum Design}()]{openquantumdesigna}
Open Quantum Design. \url{https://openquantumdesign.org/} (accessed 2025-05-06).

\bibitem[{Michael P A Murphy}(2024{\natexlab{a}})]{michaelpamurphy2024}
M. P. A. Murphy. Canada’s Alliance Politics and the Revolution in Quantum Military Affairs. Dec. 2024. \url{https://www.cigionline.org/publications/canadas-alliance-politics-and-the-revolution-in-quantum-military-affairs/} (accessed 2025-06-16).

\bibitem[{Michael P A Murphy}(2024{\natexlab{b}})]{michaelpamurphy2024a}
M. P. A. Murphy. Digital Ethics, Gender-Based Analysis and Canada’s Quantum Strategy. Aug. 2024. \url{https://www.cigionline.org/publications/digital-ethics-gender-based-analysis-and-canadas-quantum-strategy/} (accessed 2025-06-16).

\bibitem[{Michael P A Murphy}(2025)]{michaelpamurphy2025}
M. P. A. Murphy. Canada as a Norm Entrepreneur in Quantum Science and Technology. Mar. 2025. \url{https://www.cigionline.org/publications/canada-as-a-norm-entrepreneur-in-quantum-science-and-technology/} (accessed 2025-06-16).

\bibitem[{Kristen Csenkey}(2025)]{kristencsenkey2025}
K. Csenkey. Governing the Risks of Quantum-Enhanced Transportation Systems. Feb. 2025. \url{https://www.cigionline.org/publications/governing-the-risks-of-quantum-enhanced-transportation-systems/} (accessed 2025-06-16).

\bibitem[{Think 7}({\natexlab{a}})]{think7}
Think 7. Transformative Technologies — AI and Quantum. \url{https://www.think7.org/task-forces/transformative-technologies-ai-and-quantum/} (accessed 2025-06-10).

\bibitem[{Think 7}({\natexlab{b}})]{think7a}
Think 7. Enabling Quantum Technology Cooperation: A Strategic Priority for the G7 Ecosystem in the Global Race. \url{https://www.think7.org/publications/enabling-quantum-technology-cooperation-a-strategic-priority-for-the-g7-ecosystem-in-the-global-race/} (accessed 2025-04-23).

\bibitem[Seskir et~al.(2023)Seskir, Umbrello, Coenen, and Vermaas]{seskir2023}
Z. C. Seskir et al., Democratization of Quantum Technologies, Quantum Sci. Technol., 8, 024005, 2023.

\bibitem[Chakraborty et~al.(2025)Chakraborty, {de Laat}, and Laflamme]{chakraborty2025}
R. Chakraborty, K. de Laat, and R. Laflamme, Canada’s Migration to Post-Quantum Cryptography: Evaluating the Strategic Leadership Roles of Public and Private Sectors, In preparation, 2025.

\end{thebibliography}

\end{document}